\documentclass[sigconf]{acmart}

\AtBeginDocument{%
  }

\setcopyright{acmlicensed}
\copyrightyear{2025}
\acmYear{2025}
\acmDOI{XXXXXXX.XXXXXXX}

\acmConference[HT '25]{}{September 15-19, 2025}{Chicago, IL}
\acmISBN{978-1-4503-XXXX-X/18/06}

\graphicspath{{./images/}} 

\copyrightyear{2025}
\acmYear{2025}
\setcopyright{rightsretained}
\acmConference[HT Adjunct 2025]{Adjunct Proceedings of the 36th ACM Conference on Hypertext and Social Media}{September 15--18, 2025}{Chicago, IL, USA}
\acmBooktitle{Adjunct Proceedings of the 36th ACM Conference on Hypertext and Social Media (HT Adjunct 2025), September 15--18, 2025, Chicago, IL, USA}
\acmDOI{10.1145/3720533.3750065}
\acmISBN{979-8-4007-1533-4/2025/09}

\begin{document}

\title{Agency Among Agents}
\subtitle{Designing with Hypertextual Friction in the Algorithmic Web}

\author{Sophia Liu}
\email{sophiawliu@berkeley.edu}
\orcid{0009-0008-7746-0749}
\affiliation{%
  \institution{University of California, Berkeley}
  \city{Berkeley}
  \state{CA}
  \country{USA}}
  
\author{Shm Garanganao Almeda}
\email{shm.almeda@berkeley.edu}
\orcid{0000-0001-7660-313X}
\affiliation{%
  \institution{University of California, Berkeley}
  \city{Berkeley}
  \state{CA}
  \country{USA}
}

\renewcommand{\shortauthors}{Liu et al.}

\begin{abstract}
Today's algorithm-driven interfaces, from recommendation feeds to GenAI tools, often prioritize engagement and efficiency at the expense of user agency. As systems take on more decision-making, users have less control over what they see and how meaning or relationships between content are constructed. This paper introduces "Hypertextual Friction," a conceptual design stance that repositions classical hypertext principles---friction, traceability, and structure---as actionable values for reclaiming agency in algorithmically mediated environments. Through a comparative analysis of real-world interfaces---Wikipedia vs. Instagram Explore, and Are.na vs. GenAI image tools---we examine how different systems structure user experience, navigation, and authorship. We show that hypertext systems emphasize provenance, associative thinking, and user-driven meaning-making, while algorithmic systems tend to obscure process and flatten participation. We contribute: (1) a comparative analysis of how interface structures shape agency in user-driven versus agent-driven systems, and (2) a conceptual stance that offers hypertextual values as design commitments for reclaiming agency in an increasingly algorithmic web.
\end{abstract}

%%
%% The code below is generated by the tool at: http://dl.acm.org/ccs.cfm




\maketitle

\section{\textbf{Introduction}}
The web has become increasingly agentic---and increasingly opaque. Interactions with digital information, once defined by human-authored trails, are now dominated by algorithmic feeds and generative systems that anticipate user needs before intent is even articulated \cite{10.1145/3589335.3641240}. This shift reflects more than a change in interface design; it signals a broader transformation in the structure of the web itself---one that enacts a technocapitalist worldview of efficiency, optimization, and seamlessness \cite{oas2024hyperoptimization}.

We draw from prior work that reimagines friction not as a flaw but as a meaningful affordance in interaction design \cite{10.1145/2702123.2702547} \cite{kreminski2021reflective}---and offer a provocation: What if we revisited the friction of hypertext, not with nostalgia, but as a speculative framework for reclaiming agency?

Here, agency means more than the ability to click or customize. It is the capacity to navigate, trace, and compose meaning across a digital landscape. Where algorithmic systems guide users down invisible, pre-optimized paths \cite{10.1145/3374218}, hypertextual systems foreground connection and deliberateness.

For Engelbart and Nelson, hypertext assumed knowledge is constructed---not delivered---and that navigation \textit{is} authorship \cite{amisola2024becoming}. Today's web reflects a different logic: platformized, ranked, and automated \cite{platformization}. Even generative tools like DALL·E collapse user input into one-click outputs \cite{10.1145/3613904.3642466}, displacing exploration with frictionless production.

Drawing on this lineage, we ask: how might hypertext-driven interactions deliberately reintroduce friction into systems designed to eliminate it?

This Blue Skies paper contributes:

\begin{enumerate}
    \item A comparative analysis of four real-world systems---Wikipedia, Instagram Explore, Are.na, and DALL·E---showing how interface design shapes authorship, navigation, and meaning-making.
    \item A conceptual design stance---"Hypertextual Friction"---that treats \textbf{friction}, \textbf{traceability}, and \textbf{structure} as values for reclaiming agency in hybrid systems, favoring interpretation over prescription and active participation over passive use.
\end{enumerate}

We argue for systems that foreground deliberation, provenance, and user authorship---where meaning is composed, not delivered, and users are co-authors, not just subjects of computation.

\section{\textbf{Related Work}}
There is growing concern across HCI and hypertext communities about the impact of algorithmic systems on user agency. Eslami et al. \cite{10.1145/2702123.2702556} showed that most users were unaware of algorithmic filtering in feeds, and that revealing this process altered their sense of control. Alvarado and Waern \cite{10.1145/3173574.3173860} formalized this concern through the Algorithmic Experience framework, which articulates how users experience algorithmic mediation and identifies key dimensions of agency---including transparency, feedback, and user understanding. Their framework underscores that algorithms shape not only what users see, but how they perceive and relate to the systems they interact with.

Amid trends toward automation and seamless interaction, friction has emerged as a productive counter-strategy. From Hallnäs and Redström's foundational work on \textit{Slow Technology} \cite{10.1007/PL00000019} to more recent explorations \cite{10.1145/2317956.2318088} \cite{seamful}, researchers have reframed friction and slowness not as a usability flaw, but as a feature of design with distinct affordances---finding that resisting ease or efficiency can deepen engagement and add depth and texture to interactions.

A number of recent works in the hypertext community have advocated for the development of new conceptual frameworks and interventions to respond to the impact of Generative AI systems. Atzenbeck et al.'s "Unwinding AI's Moral Maze" \cite{10.1145/3648188.3678213} critiques GenAI systems as ethically ambiguous "moral mazes"---making a conceptual argument for hypertextual values like provenance, grounded in the community's long-standing commitment to traceable, user-authored structures. More concretely, Roßner et al.'s SPORE system \cite{10.1145/3603163.3609075} operationalizes these values through a spatial hypertext interface for narrative co-creation. SPORE combines visual layout, recommendation, and generative tools to support "storybreaking" as a collaborative, interpretive process. Informal deployments revealed that users gravitated toward its associative layout and adapted its friction as a form of creative play. These findings suggest friction and structure can be compelling in practice.

Beyond technical frameworks, artists and theorists have revisited hypertext as a lens for authorship and agency. In their performance lecture "Becoming Hypertext," artist Chia Amisola explores hyperlinking as a mode of personal and epistemological authorship---arguing that "readers are also authors," and that "linking is also authorship" \cite{amisola2024becoming}. Their work reimagines hypertext not just as a technical structure, but as an expressive, embodied form of navigating and composing meaning.

Our work builds on this momentum by advocating for a design stance we call "Hypertextual Friction"---not because hypertext and friction are synonymous, but because the structures of hypertext---visible connections, associative trails, deliberate navigation---create the kinds of friction that support agency. Whereas SPORE demonstrates the promise of spatial hypertext in one domain, we extract and generalize its underlying design values across interface types. While prior systems gesture toward interpretability \cite{10.1145/3313831.3376219} or modularity \cite{golechha2024trainingneuralnetworksmodularity}, we position hypertext as a structural intervention---one that counters algorithmic acceleration with form.

This stance aligns with recent work framing hypertext not merely as a technical structure, but as a method or philosophy of interaction grounded in post-structuralist and existentialist values \cite{10.1145/3342220.3343669} \cite{10.1145/3603163.3609048}. Our contribution builds on this framing by offering a conceptual stance that is also actionable---inviting systems that foreground reflection, authorship, and user agency in algorithmic contexts.

\section{\textbf{Interface Paradigms}}
We frame our analysis through two dominant paradigms ~\footnote{While we present these as contrasting paradigms, many real-world systems---like DALL·E’s chain-of-thought prompting or Instagram’s feed personalization---inhabit increasingly hybrid spaces that blend automation with user authorship, further underscoring the need for frameworks that can guide design across this evolving spectrum.}: algorithmic systems and hypertextual systems. Both facilitate digital exploration, but differ in how they structure information, guide navigation, and position the user.

\subsection{\textbf{Algorithmic Systems}}
While all computational systems rely on algorithms in a technical sense, we use the term "algorithmic systems" here to refer specifically to an interaction design paradigm that centers predictive, generative, or adaptive outputs driven by computational models or agents \cite{roy2022recommender}. These include recommendation engines (e.g., Spotify's song recommendations), feeds curated by recommendation engines (e.g., TikTok's For You page), and agentic assistants (e.g., ChatGPT \cite{openai2023chatgpt}, DALL·E \cite{DBLP:journals/corr/abs-2102-12092}, and GitHub Copilot \cite{github2021copilot}).

Such systems streamline experiences, removing the need for explicit navigation \cite{roy2022recommender}. Content is surfaced, suggested, or produced automatically, shaping interaction around prediction and immediacy. In many cases, the underlying logic, selection process, or data provenance is hidden from view \cite{rudin2019stop} \cite{HOSSEINZADEHAGHDAM201989}.

\subsection{\textbf{Hypertextual Systems}}
We use the term "hypertextual systems" to describe an interaction paradigm that, in contrast, structures information through visible, intentionally authored links. This includes formats like personal blogs and forums, as well as systems like Wikipedia \cite{wiki:xxx}, Obsidian \cite{obsidian}, a personal knowledge base and notetaking platform, and Are.na \cite{arena_about}, a collaborative tool for collecting, organizing, and sharing content.

They emphasize manual exploration and associative thinking. Users trace connections, follow authored trails, and construct meaning through linked structure \cite{10.1109/2.222119}. While less optimized for speed, they foreground provenance, deliberateness, and visible structure in interaction.

This framing provides the basis for our subsequent comparison of four real-world systems: two hypertextual and two algorithmic.

\section{\textbf{Comparative Interface Analysis}}
To ground our analysis, we examine two interface pairings: one between information systems (Wikipedia vs. Instagram Explore), and one between creative tools (Are.na vs. DALL·E). We analyze how they differ in the ways that they structure navigation, surface content, and position the user---as agent or input. These comparisons reveal how hypertextual vs. algorithm-driven logics shape authorship, interpretation, and experience.
\begin{figure*}[h]
  \centering
  \begin{minipage}[t]{0.32\textwidth}
    \centering
    \includegraphics[height=5cm]{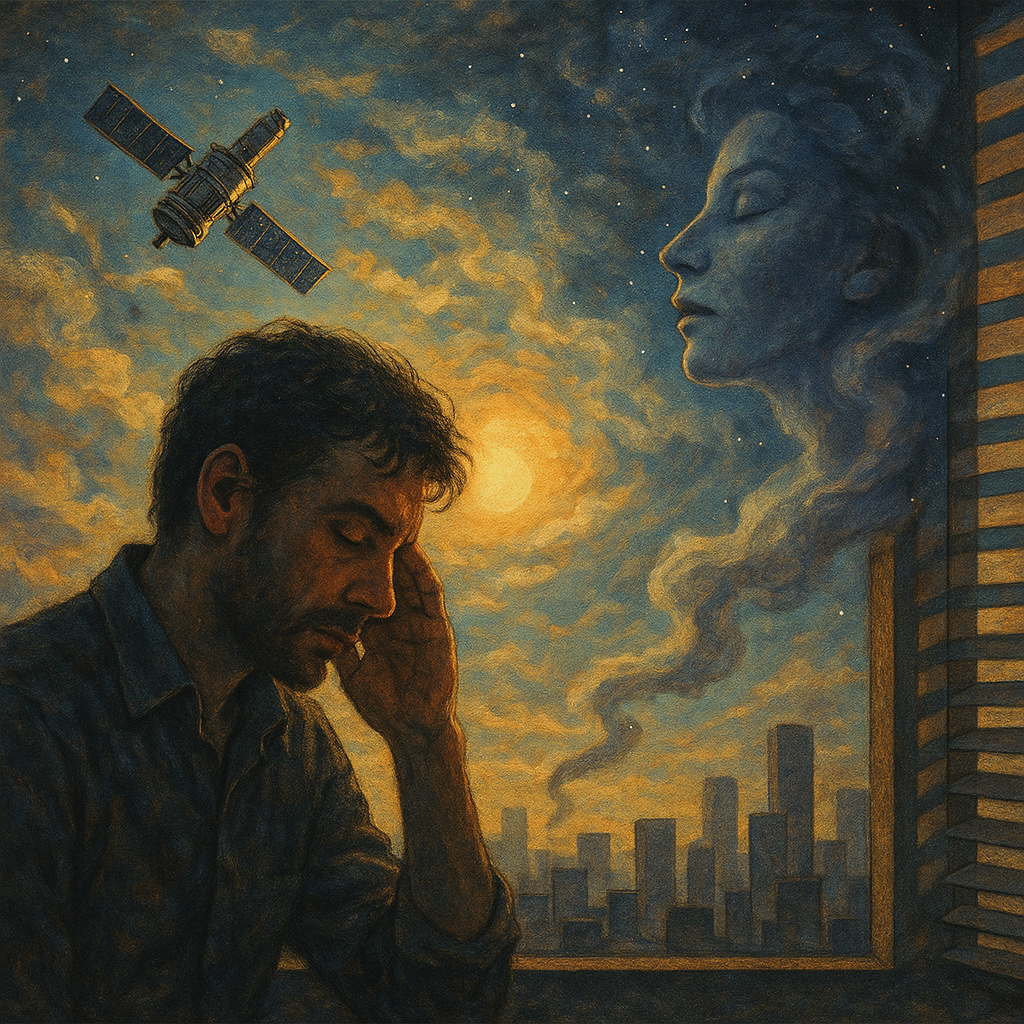}
    
    \vspace{2pt}
    \small DALL·E output (1): literal lyrics-to-image prompt
  \end{minipage}
  \hfill
  \begin{minipage}[t]{0.32\textwidth}
    \centering
    \includegraphics[height=5cm]{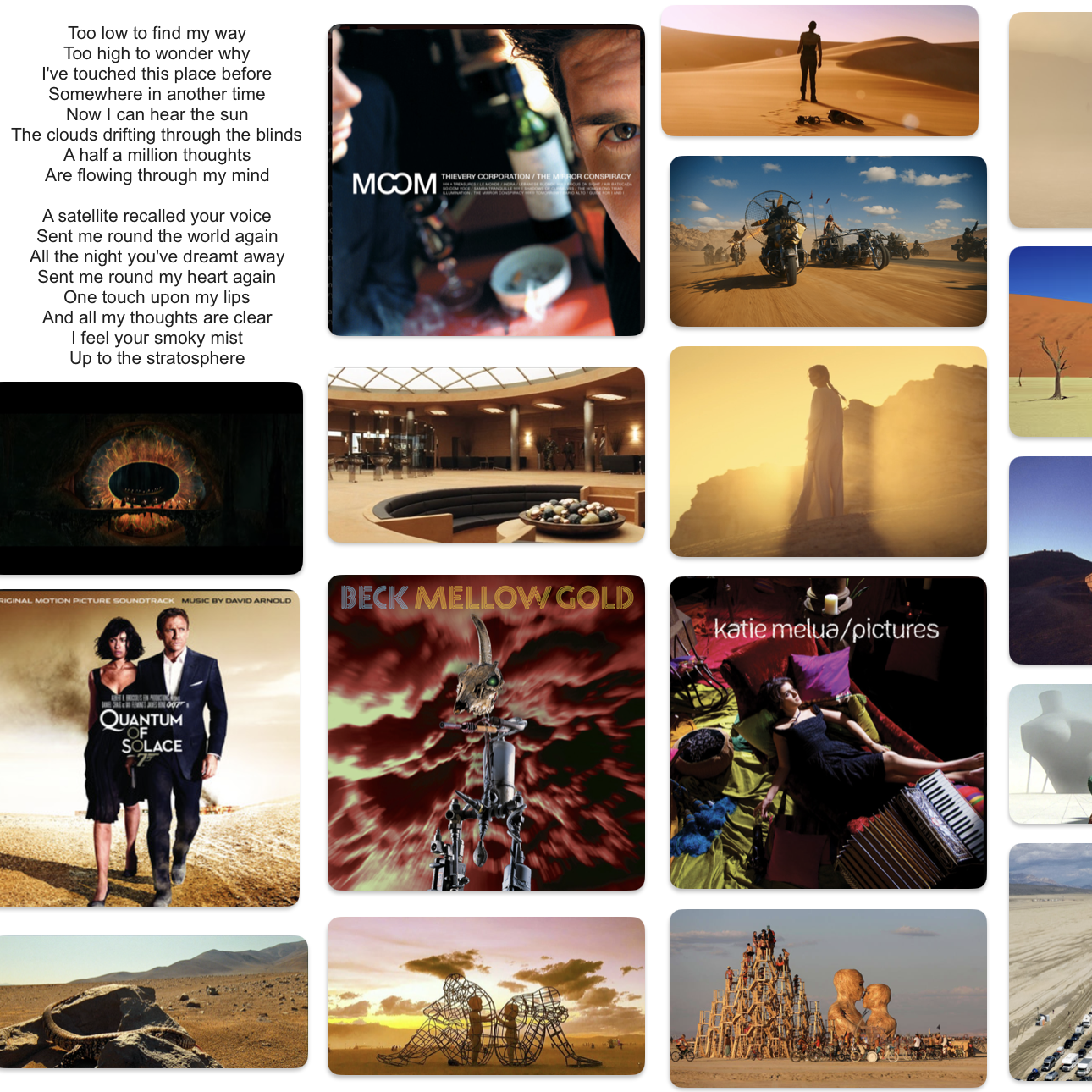}
    
    \vspace{2pt}
    \small Manual curation: personalized moodboard assembled via associative browsing and hybrid tools
  \end{minipage}
  \hfill
  \begin{minipage}[t]{0.32\textwidth}
    \centering
    \includegraphics[height=5cm]{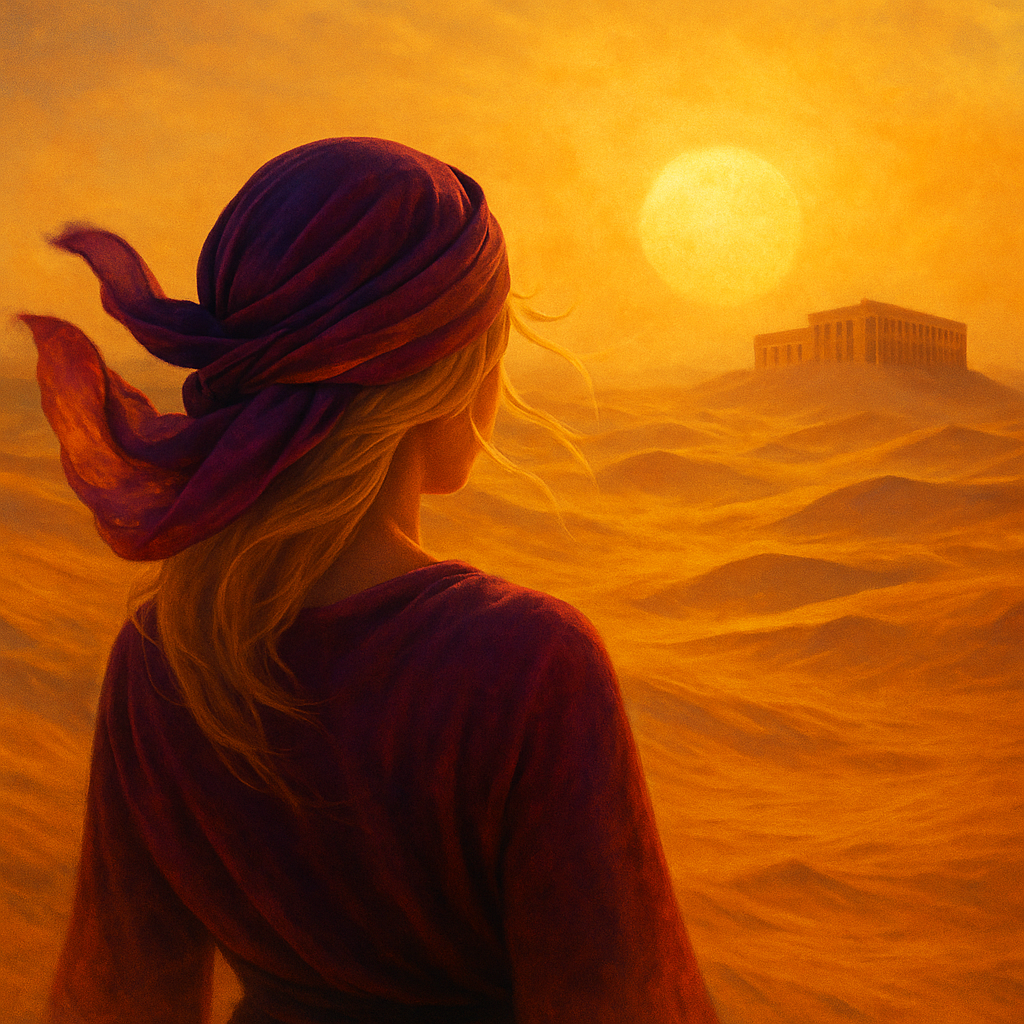}
    
    \vspace{2pt}
    \small DALL·E output (2): refined prompt shaped by moodboard and hypertextual exploration
  \end{minipage}
  
\caption{Three approaches to visualizing the "vibe" of Thievery Corporation’s "Lebanese Blonde." Left: A literal interpretation generated by DALL·E from a lyrics-only prompt. Middle: A deeply personal, hybrid moodboard created through associative, exploratory curation across Are.na, Pinterest, Wikipedia, and media memory. Right: A more resonant DALL·E output informed by the curated reference set and structured, interpretive prompting. The series illustrates how intentional authorship and hypertextual scaffolding can reclaim personal resonance and interpretive depth in generative workflows.}

  \label{fig:lebanese-blonde-comparison}
\end{figure*}

\subsection{\textbf{Wikipedia vs. Instagram Explore}}
\textbf{Wikipedia and the Hypertextual Path}
Wikipedia invites users to navigate a web of visible links that reward curiosity and associative thinking. Far from a static resource, it functions as a hypertext system that encourages open-ended, exploratory browsing. According to the Wikimedia Foundation, about one-third of sessions begin with a topical query but evolve into unstructured discovery ~\cite{allemandou2018wikistats}.

A reader might follow a trail like \textit{Hypertext} → \textit{Ted Nelson} → \textit{Xanadu} → \textit{Electric Light Orchestra}, crossing theory, history, and pop culture. Each click is a decision---what Amisola might describe as an act of authorship in the shaping of a personalized journey \cite{amisola2024becoming}. We propose that meaning emerges through juxtaposition; discovery is built, not served.

Wikipedia fosters what researchers have called \textit{active wandering} \cite{10.1145/1149941.1149946}, enabling serendipity through its intentionally linked structure. Unlike algorithmic systems that prescribe a path, it invites users to forge their own. Ted Nelson famously called this structure "intertwingularity"---the overlapping, nonlinear mesh of ideas that defies simple hierarchy \cite{nelson1974computer}. Even the interface reflects this epistemology: hyperlinks are not just navigational, but meaning-making cues \cite{recentarts2022hypertext}\cite{nelson1974computer}.

This slowness is not a flaw but a feature. By resisting prediction, Wikipedia enables interpretive agency. It doesn’t guess what the user wants---it invites the user to find it.

\textbf{Instagram Explore and the Algorithmic Stream}
Instagram’s Explore page exemplifies algorithmic interaction without paths. The interface presents a continuous stream of content, optimized for attention through reactive design. Its ranking pipeline weighs engagement metrics like likes, shares, view duration, interaction history, and media virality to prioritize high-arousal, visually striking content \cite{giraldo2020attention} \cite{HOSSEINZADEHAGHDAM201989} \cite{roy2022recommender}. The logic behind this curation is opaque: there is no visible trail, no provenance---only a stream that refreshes ad infinitum.

While the interface appears to offer discovery, the experience is tightly personalized. The recommendation system narrows what is shown based on a user’s past behaviors, presenting content that mirrors prior interactions and reinforces existing patterns \cite{KOLAJO2020102348} \cite{roy2022recommender}. Users do not navigate across ideas; they scroll. In place of authored structure, Instagram offers associative disassociation---content stitched together by unseen algorithms rather than intentional links.

Studies liken this experience to "digital junkfood" \cite{10.1145/3648188.3678163}: high-engagement, low-context media that fragments attention and limits depth. Curation becomes invisible and decision-making outsourced. Rather than constructing meaning, users are fed content.

\subsection{\textbf{Are.na vs. DALL·E}}
\textbf{Are.na and the Pleasure of Curation}
Are.na \cite{arena_about} is a visual organization and research platform used by designers, artists, and researchers to collect, arrange, and annotate media. Users build collections by linking "blocks"---images, texts, and URLs---into spatial, non-hierarchical threads. These blocks accumulate meaning through their placement and relation, rather than algorithmic ranking. The platform has no likes, followers, or feeds; it prioritizes intentionality over visibility.

In this context, "vibe" is not generated, but constructed---through manual arrangement, contextual layering, and associative logic. Are.na exemplifies hypertextual authorship: traceable and emergent. Its interface mirrors the logic of \textit{intertwingularity}, enabling nonlinear thinking and meaning-making through a mesh of visible, overlapping connections \cite{arena_about}. Like a digital sketchbook or garden, Are.na enables world-building through trails of thought.

\begin{table*}[t]
\centering
\small
\setlength{\tabcolsep}{4pt} % Reduce space between columns
\renewcommand{\arraystretch}{1.2} % Optional for spacing
\caption{Values of friction, traceability, and structure that differentiate hypertextual (user-driven) and algorithmic (agent-driven) systems. These values underpin our design stance for agency in hybrid interfaces.}
\label{tab:interaction-duality}
\begin{tabular}{|p{3.2cm}|p{4.2cm}|p{4.2cm}|p{4.2cm}|}
\hline
\textbf{System Type} & \textbf{Friction} & \textbf{Traceability} & \textbf{Structure}\\
\hline
\textbf{Hypertextual (User-Driven)}& Slower, intentional navigation (e.g., manual linking, decision points) & Visible sources, contextual trails, historical lineage & Meaning constructed through linking, annotation, and arrangement \\
\hline
\textbf{Algorithmic (Agent-Driven)}& Seamless, predictive interaction (e.g., feed scroll, prompting) & Opaque logic, hidden influence, flattened provenance & Output-focused; little space for user-led composition \\
\hline
\end{tabular}
\end{table*}

\textbf{DALL·E and the Elimination of Process}
DALL·E is a generative image tool that transforms natural language prompts into synthetic visuals using large-scale diffusion models \cite{DBLP:journals/corr/abs-2102-12092}. It offers immediacy and polish: users input a phrase and receive a finished image within seconds. But this speed comes at a cost. The process is opaque and often flattens meaning into surface aesthetics \cite{almeda2024dreamsheets}.

To explore how generative tools shape visual world-building, we prompted DALL·E to interpret the "vibe" of Thievery Corporation’s "Lebanese Blonde" using only a lyrics-based prompt. The resulting images rendered the literal content---mist, blinds, satellites---but failed to capture the song’s affective texture: the desert dreamscapes, cinematic grain, and embodied nostalgia it evoked for us. The image aligned with denotation, not resonance---with surface, not the deeply personal visual world the song conjured.

In response, we initiated a manual curation process (see Figure~\ref{fig:lebanese-blonde-comparison})---though not one entirely free from algorithmic influence. We began with Are.na as a starting point, using it to map out associative links and paths of inquiry. We browsed Wikipedia trails, Pinterest suggestions, and personal media references. While only some of these were algorithmically shaped, together they supported a hypertextual journey---navigated with intention.

We then used Apple's Freeform \cite{apple2022freeform} to spatially arrange and layer the references into a moodboard---a flexible canvas that complemented Are.na’s link-based trailbuilding. While Are.na supported associative curation through structured connections, Freeform enabled visual synthesis. Together, they supported a process of intentional exploration and reflection---externalizing a vibe that was both deeply personal and structurally traceable.

This process became a kind of hypertextual composition \cite{amisola2024becoming}, where each image or reference led to the next. Meaning wasn’t just generated---it was curated through a frictive, slow, and intentional process. We returned to DALL·E with a screenshot of the moodboard and short personal narratives written during reflection in Freeform. The resulting image resonated more---not because the model improved, but because the input was shaped by intentional labor. Meaning emerged from the path, not the endpoint. This illustrates how GenAI systems can benefit from hypertextual scaffolding---not to optimize output, but to foreground reflection, authorship, and agency in the creative process.

\section{\textbf{Hypertextual Friction: A Design Stance}}
What would it mean to design for agency in a web increasingly shaped by agents?

From our comparative analysis, we articulate "Hypertextual Friction" as a conceptual design stance: one that centers three interface-level values---\textbf{friction}, \textbf{traceability}, and \textbf{structure}. These echo early hypertext systems like Bush’s trails \cite{bush:1945:awmt} and Nelson’s intertwingularity \cite{nelson1974computer}. We reorient these values toward contemporary algorithmic systems, which flatten decision-making and obscure provenance \cite{oas2024hyperoptimization}. "Hypertextual Friction" treats these values not as relics, but as actionable scaffolds for reclaiming agency in hybrid, automation-shaped interfaces.

\textbf{Friction} slows users to surface intention. In Wikipedia, each hyperlink requires deliberation; navigation is situated and authored. This reflects Amisola’s belief that following a trail of thought is itself an act of authorship \cite{amisola2024becoming}. By contrast, interfaces like Instagram Explore suppress decision-making in favor of uninterrupted recommendation streams. In hybrid systems, friction might take the form of visible forks in content, reflective pauses, or points of editorial choice---gestures that reintroduce agency into the flow.

\textbf{Traceability} makes visible how information is sourced and sequenced. Hypertext foregrounds provenance---Wikipedia’s citations and Are.na’s sourced blocks embody this ethic. GenAI systems like DALL·E often collapse the path from prompt to output, erasing history and influence. Traceability shows how something was made---not just what. Hybrid tools could show source trails and remix histories---letting users retrace or repurpose what they find.

\textbf{Structure} gives users a scaffold for composing and relating ideas. Unlike prompt-response cycles, hypertext supports nonlinearity and reuse. Are.na collections show how structure emerges through juxtaposition, context, and associative logic. Hybrid systems could extend this by embedding generative outputs into editable canvases or semantic trails---where meaning evolves through interaction.

These values position hypertext as a foundation for future systems. We advocate for legible, associative interfaces that embed agency. "Hypertextual Friction" outlines a path toward hybrid tools that support meaningful authorship.

\section{\textbf{Conclusion}}
In an algorithmic web increasingly populated by feeds, recommendation engines, and generative assistants, we argue for a renewed commitment to user agency. While prior work has critiqued algorithmic opacity and called for transparency, we suggest this is not enough. Agency must be reclaimed not only through explanation, but through the structure of interaction itself.

We introduce "Hypertextual Friction"---a stance that treats friction, traceability, and structure as actionable interface values. Rather than nostalgia, hypertext offers a timely framework for resisting the seamlessness and flattening of algorithmic systems. Comparative analysis of real-world interfaces shows how interaction structures shape how users navigate, interpret, and author meaning.

We contribute a vocabulary that invites designers to rethink system design and embed agency at the interface level. Future empirical work should examine how these values shape user experience and scale in hybrid systems.

\textit{Agency among agents} requires more than good prompt engineering. It calls for structural alternatives. In this light, hypertext offers more than critique---it offers a design path forward.

\begin{acks}

In line with this paper’s themes, OpenAI’s ChatGPT was used as a tool for editing and synthesis, with the aim of supporting human reflection rather than frictionless generation.

\end{acks}

\bibliographystyle{ACM-Reference-Format}
\bibliography{references.bib}

\end{document}